\documentclass[preprint, pra, aps,floatfix, showpacs,amsmath]{revtex4}
\usepackage{epsfig}
\begin{document} 
\newcommand{\newc}{\newcommand}
\newc{\beq}{\begin{equation}}
\newc{\eeq}{\end{equation}}
\newc{\kt}{\rangle}
\newc{\br}{\langle}
\newc{\beqa}{\begin{eqnarray}}
\newc{\eeqa}{\end{eqnarray}}
\newc{\pr}{\prime}
\newc{\longra}{\longrightarrow}
\newc{\ot}{\otimes}
\newc{\rarrow}{\rightarrow}
\newc{\h}{\hat}
\newc{\bom}{\boldmath}
\newc{\btd}{\bigtriangledown}
\newc{\al}{\alpha}
\newc{\be}{\beta}
\newc{\ld}{\lambda}
\newc{\sg}{\sigma}
\newc{\p}{\psi}
\newc{\eps}{\epsilon}
\newc{\om}{\omega}
\newc{\mb}{\mbox}
\newc{\tm}{\times}
\newc{\hu}{\hat{u}}
\newc{\hv}{\hat{v}}
\newc{\hk}{\hat{K}}
\newc{\ra}{\rightarrow}
\newc{\non}{\nonumber}
\newc{\ul}{\underline}
\newc{\hs}{\hspace}
\newc{\longla}{\longleftarrow}
\newc{\ts}{\textstyle}
\newc{\f}{\frac}
\newc{\df}{\dfrac}
\newc{\ovl}{\overline}
\newc{\bc}{\begin{center}}
\newc{\ec}{\end{center}}
\newc{\dg}{\dagger}
\newc{\prh}{\mbox{PR}_H}
\newc{\prq}{\mbox{PR}_q}
\newc{\tr}{\mbox{Tr}}

\title{Fluctuations of finite-time stability exponents in the standard map and the detection of small islands }
\author{Steven Tomsovic\footnote{Martin Gutzwiller Fellow, 2006/2007; permanent address: Department of Physics and Astronomy, Washington State University, Pullman, WA  99164-2814}}
\author{ Arul Lakshminarayan\footnote{Permanent address: Department of Physics, Indian Institute of Technology Madras, Chennai, 600036, India.}}
\affiliation{Max-Planck-Institut f\"ur Physik komplexer Systeme, N\"othnitzer Stra$\beta$e 38, D-01187 Dresden, Germany}
\date{\today}
\begin{abstract}

Some statistical properties of finite-time stability exponents in the standard map can be estimated analytically.  The mean exponent averaged over the entire phase space behaves quite differently from all the other cumulants.  Whereas the mean carries information about the strength of the interaction, and only indirect information about dynamical correlations, the higher cumulants carry information about dynamical correlations and essentially no information about the interaction strength.  In particular, the variance and higher cumulants of the exponent are very sensitive to dynamical correlations and easily detect the presence of very small islands of regular motion via their anomalous time-scalings.  The average of the stability matrix' inverse trace is even more sensitive to the presence of  small islands and has a seemingly fractal behavior in the standard map parameter.  The usual accelerator modes and the small islands created through double saddle node bifurcations, which come halfway between the positions in interaction strength of the usual accelerator modes, are clearly visible in the variance, whose time scaling is capable of detecting the presence of islands as small as $0.01\%$ of the phase space.  We study these quantities with a local approximation to the trace of the stability matrix which significantly simplifies the numerical calculations as well as allows for generalization of these methods to higher dimensions. We also discuss the nature of this local approximation in some detail.
\end{abstract}

\maketitle

\section{Introduction}
\label{intro}

A central and widely used measure of instability in dynamical systems is the Lyapunov exponent \cite{Lichtenberg92,Ott97}.  If a positive Lyapunov exponent is found for a continuous family of trajectories, then it is taken as a  measure of their dynamical chaos.  It is natural that the Lyapunov exponent, which is defined as a time-approaching-infinity limiting value, when calculated numerically, would be approximated by considering finite time segments of the infinite history of a trajectory.  This has led to the study of the so-called finite time or local Lyapunov exponents (FTLE) \cite{Ott97} in a variety of 
contexts.  For example, FTLE in one-dimensional maps \cite{Fujisaka83,Theiler95,Prasad99,Anteneodo04}, Hamiltonian systems \cite{Sepulveda89,Amitrano92,Amitrano93}, advection and turbulent flows \cite{Beigie93,Lapeyre02,Arratia05}, form just a part of a rather large literature. Additionally in most experimental situations one encounters finite time quantities.   Such FTLE have been the subject of numerous studies  and have been  experimentally measured in many contexts such as for instance in the study of passively advected particles \cite{Arratia05}.  There is also a very intimate relationship between the theory, methods, and measures found in the subject of Anderson localization\cite{Crisanti93} and FTLE.  Although, typically dynamical correlations are assumed vanishing or if correlations are accounted for\cite{Rechester80}, then they are not the kind which is found in Hamiltonian flows where all the complexity of boundaries between chaotic and regular motion may exist. In other words these correlations are then calculated assuming that a typical trajectory covers phase space uniformly, whereas the presence of islands, however small, ensures that this is not true.

A very closely related, but less directly studied, quantity is the stability exponent which is defined via the eigenvalues of the stability matrix.  A detailed study of this measure vis-a-vis the Lyapunov exponent was carried out in \cite{Goldhirsch87} where it was conjectured that the Lyapunov and stability exponents were equal under a broad range of dynamical assumptions. It turns out that some of the studies that purported to be about the FTLE have actually been studies of the stability exponents instead.  In semiclassical mechanics, such as in the Gutzwiller trace formula \cite{Gutzwiller90},  one of the natural classical quantities that appears is, in fact,  the stability exponent and not the Lyapunov exponent.  In higher dimensional systems (larger than two-degrees-of-freedom), it is the sum of the positive stability exponents that appears and in the infinite-time limit this tends  to the Kolmogorov-Sinai (K-S) entropy. Since it  has long been appreciated that the small wavelength limit and the large time limit in quantum mechanics are not interchangeable, we may expect that for fixed wavelength, it is the finite-time stability exponents, to be defined below more precisely, which may be relevant; see for example \cite{Silvestrov03}. In fact, many of the interesting ways in which the fluctuations in the stability exponents might potentially show up in quantum mechanics and semiclassical theories remains to be investigated. This forms our primary motivation for studying the stability exponents, their distribution and their scaling properties. Nevertheless, in this paper we restrict our attention to purely classical and low-dimensional systems (two-dimensional area preserving maps) to establish basic properties and approximations, as well as to demonstrate their usefulness in detecting small stable dynamical structures. Naturally there are many parallels in the large literature on FTLE both of a quantitative and qualitative character given that for a broad class of dynamical systems the infinite-time stability exponents are identical to the Lyapunov exponent.  Our  results presented here generally for stability exponents are thus of relevance to the literature on the FTLE as well. 

In the following section we first introduce the model in use throughout (the standard map) and an approximation for the stability exponent which provides a practical, calculational simplification that can be generalized to higher dimensions.   Its ergodic value has corrections to the well-known formula of Chirikov \cite{Chirikov79}.  In section \ref{average}, the local properties of the stability exponents are studied. We image the finite-time stability exponents in phase space and show their close relationship to the stable and unstable manifolds of the map, and show their connection to locating small non-hyperbolic regions.  In section \ref{fluc}, the fluctuations of the stability exponents are investigated, and a formula for the ergodic average of the variance is derived.  The scaling of the variance is shown to be a sensitive sensor for small islands.  It is also shown that quantities that incorporate higher cumulants and that are of natural semiclassical relevance are even better sensors of such phase-space structures.  In the appendices, formulas are collected for an expansion of the exact trace of the stability matrix as well as results for the time scaling of the stability exponent higher cumulants.

\section{Background, an approximation, and ergodic average}
\label{class}
In this section we introduce the kicked rotor, provide definitions of the quantities of interest, explore an approximation of great practical utility, and use it to generate an improved analytic estimate of the Lyapunov exponent for the standard map.
\subsection{The Hamiltonian}
\label{backg}
Consider the 2-D standard map \cite{Lichtenberg92,Ott97} as a well-suited vehicle for our investigations.   The methods and conclusions we reach will have wider applicability as this map represents the generic, local dynamics of any integrable system with shear in action-angle coordinates and shows how that structure gets altered by perturbation.   A general kicked rotor is a mechanical-type particle constrained to move on a ring that is kicked instantaneously every multiple of a unit time, $t=n\tau$.  Supposing the radius of the ring to be $1/2\pi$ and $\tau=1$, the Hamiltonian takes the form
\begin{equation}
\label{krg}
H(q,p) = {p^2\over 2 } + V(q) \sum_{n=-\infty}^\infty \delta(t-n).
\end{equation}
where $V(q)$ is a function periodic on the interval $q\in [0,1)$.  From $H(q,p)$, mapping equations can be given:
\begin{eqnarray}
\label{kreq}
p_{i+1} &=&  p_i -V^\prime(q_i)  \qquad {\rm\  mod\ } 1 \nonumber \\
q_{i+1} &=& q_i + p_{i+1}  \qquad\ \ \  {\rm\  mod\ } 1.
\end{eqnarray}
where we have made the choice to apply the potential kick before the free motion.  The notation $V^\prime$ indicates the derivative of $V$ with respect to $q$.  We have also restricted the phase space to the unit torus ($q,p\in [0,1)$) for convenience; this is of no consequence for the results presented in this paper.
The simplest periodic function on a ring is just the lowest harmonic and leads to the standard map. 
\begin{equation}
\label{potential}
V(q) =  -\dfrac{K}{ 4\pi^2 }\cos \left(2\pi  q \right) 
\end{equation}
Many results have long been known for the standard map\cite{Chirikov79}. At $K=0$ the map is integrable and is essentially a stroboscopic map of a freely rotating particle.  There are both rational and irrational tori, depending on whether the frequency of rotation is commensurate or not with the frequency of the strobe.  As the kicking strength is increased from zero with exactly the same frequency of stroboscopic observation, the incommensurate tori (irrational) survive small perturbations in accordance with the KAM theorem, while the commensurate (rational) ones break up into a pair of stable and unstable orbits in accordance with the Poincare -Birkhoff theorem. The phase space becomes mixed with stable and chaotic orbits for increasing $K$.  At around $K\approx 1$ the last rotational irrational KAM  tori breaks and this leads to global diffusion. Up to $K=4$ a stable fixed point persists. Beyond $K\approx5$ the standard map is considered to be largely chaotic, although it is also not proven to be completely chaotic for any value of $K$ as far as we know. At an infinity of  values of $K$ stable fixed points are known to appear in the $p=0$ line for the map on the torus that are accelerator modes for the map on the cylinder \cite{Lichtenberg92}. These typically occupy regions in the phase space whose areas scale as $1/K^2$ \cite{Chirikov79}. Otherwise, the Lyapunov exponent of the map, to a very good approximation (Chirikov's result),  increases with $K$ as $\ln(K/2)$.

\subsection{The finite time stability exponents}
For the map of Eq.~(\ref{kreq}), the stability matrix at time $t$ is given by
\begin{equation}
\label{stabm}
M_t(q_0,p_0)=\prod_{i=0}^{t-1} M(q_i) = \prod_{i=0}^{t-1} \left( \begin{array}{cc}
  1     &    -V^{\prime\prime}(q_i) \\
  1    &  1-  V^{\prime\prime}(q_i),
\end{array}\right)
\end{equation}
where the $M(q_i)$ are local one-step stability matrices.  The finite time Lyapunov exponents  are defined via the singular values of $M_t$, which are the eigenvalues of $M_t^T M_t$ and turn out to be non-negative real numbers. The eigenvalues of stability matrices $M_t$ on the other hand can in general be complex.  To be technically correct we refer to the  positive finite-time ``stability exponents'' (FTSE), $\lambda_t(q_0,p_0)$, where the time $t$ is the integer number of applications of the mapping equations and $(q_0,p_0)$ is the initial condition.  Consider only  initial conditions $(q_0,p_0)$ for which the orbits are unstable, in which case for $2D$ area preserving maps such as we are presently studying the eigenvalues have to be real.  The $\lambda_t(q_0,p_0)$ can be expressed in terms of the trace of the stability matrix as follows
\begin{equation}
\label{deflambda}
\lambda_t(q_0,p_0) = {1\over t} \ln \left |{\mbox{\tr}[M_t(q_0,p_0)] + \sqrt{\mbox{\tr}[M_t(q_0,p_0)]^2-4}\over 2}\right| \approx {1\over t} \ln \left| \tr[M_t(q_0,p_0)] \right| 
\end{equation}
where $\tr(...)$ denotes the trace operation.  The approximation applies for $t$ large enough.  In that case, the approximation error is exponentially small in the time except for stable trajectories. The generalization to higher dimensions is through the exponential growth rate of the magnitudes of the eigenvalues of the stability matrix (the Jacobian) at time $t$.

The natural quantity, which occurs in semiclassical expressions, is  $|\mbox{Det}(M_t-I)|$ whose large time behavior is determined by the sum of the positive stability exponents. For 2D area-preserving maps one has $|\mbox{Det}(I-M_t)|=|\mbox{Tr}(M_t)-2|$, and the maximum stability exponent  coincides with the K-S entropy in the $t\rightarrow \infty$ limit. Thus for the kicked rotor, studying  $|\mbox{Det}(I-M_t)|$, $|\mbox{Tr}(M_t)|$, or the K-S entropy is equivalent and our interest is in these quantities' (or their logarithms) finite-time fluctuations over phase space.  Therefore our study is naturally close to  the well-explored field of FTLE.  There are strong reasons to believe that in the infinite time limit the definitions of FTLE and FTSE are equivalent in the chaotic cases that we are concerned with, although non-generic  examples are known for which this is not true \cite{Goldhirsch87}.  Since we are dealing with finite time measures we wish to be technically correct and refer to FTSE, although these typically deviate from the FTLE by terms of the order of $1/t$. Recall that the FTLE can also be defined via the way initial error vectors $y_0$ grow in length: 
\beq
\mu_t(q_0,p_0;y_0)=\dfrac{1}{t} \ln\left(\df{\parallel M_t(q_0,p_0) \,y_0 \parallel}{\parallel y_0 \parallel}\right)
\eeq
 This definition of FTLE has the added complication that they will depend on the initial error vector $y_0$ as well.   However, in the infinite time limit this corresponds to the exponent coming from the singular values of the limit of $M_{t\rightarrow \infty}$.  Thus, $\mu_\infty$ is the Lyapunov exponent and it turns out not to have initial condition dependence or fluctuations.   There are reliable methods of evaluating each of these quantities \cite{Lichtenberg92,Greene87} (whether FTLE or FTSE) and they give essentially the same information, especially regarding the fluctuations of the exponents that we are primarily concerned with here. Higher dimensional generalizations, which are currently under study \cite{Tomsovicinprep}, involve the fluctuations of $|\mbox{Det}(M-I)|$ and therefore we prefer to concentrate on FTSE in this paper on 2D maps. 

\subsection{Approximating the trace}

Before turning to a more detailed analysis of $\lambda_t(q_0,p_0)$, its average, variance and higher order moments and cumulants, we investigate approximations to the trace.  These approximations make the trace calculations both efficient and simple, and extend to higher-dimensional systems, where they  provide significant advantages. For any stability matrix of the form of Eq.~(\ref{stabm}), it turns out that $\tr[M_t(q_0,p_0)]$ can be expressed in an exact, closed form of products of the individual traces $\tr[M(q_i)]=2-V^{\prime\prime}(q_i)$; the full expressions to all orders are found in Appendix A.  We display here the zeroth order approximation, indicated by $j=0$, and its first order correction, indicated by $j=0+1$:
\begin{equation}
\label{exacttrj0}
\tr[M_t(q_0,p_0)]_{j=0} =  \prod_{i=0}^{t-1} \left[2- V^{\prime\prime}(q_i)\right]
\end{equation}
\begin{equation}
\label{exacttrj01}
\tr[M_t(q_0,p_0)]_{j=0+1} =  \prod_{i=0}^{t-1} \left[2- V^{\prime\prime}(q_i)\right] \times \left(1 - \sum_i^{cyclic} {1\over  \left[2- V^{\prime\prime}(q_i)\right] \left[2- V^{\prime\prime}(q_{i+1})\right] }  \right)
\end{equation}
These expressions represent well the large-$K$ behavior of the exact trace $\tr(M_t(q_0,p_0))$, even though, a priori, one might suspect that estimating this trace by Eq.~(\ref{exacttrj0}) is rather inadequate.  After all, consider the fact that the leading correction, shown in Eq.~(\ref{exacttrj01}), contains $t$ terms of inverse products of pseudo-random factors which have the possibility of nearly vanishing.  Crudely speaking, as multiple trajectories are considered or as $t$ increases for an individual trajectory, it becomes more and more likely that a nearly vanishing factor will be encountered thus creating a correction term that dominates the leading term.  Worse, all the successive corrections involve combinations of the same products of inverse factors (Appendix A).  It is possible that no particular correction term dominates either. 

Fortunately and oddly enough, Eq.~(\ref{exacttrj0}) (the $j=0$ term) is an excellent approximation, and in fact, is more than adequate for analytically estimating the leading asymptotic behaviors of the finite time stability exponents.  It is even good enough to capture the power-laws  for both local averages
and variances of FTSE, which we investigate ahead. 
\begin{figure}[t]
\begin{center} 
\leavevmode 
\epsfxsize = 16.0cm 
\epsfbox{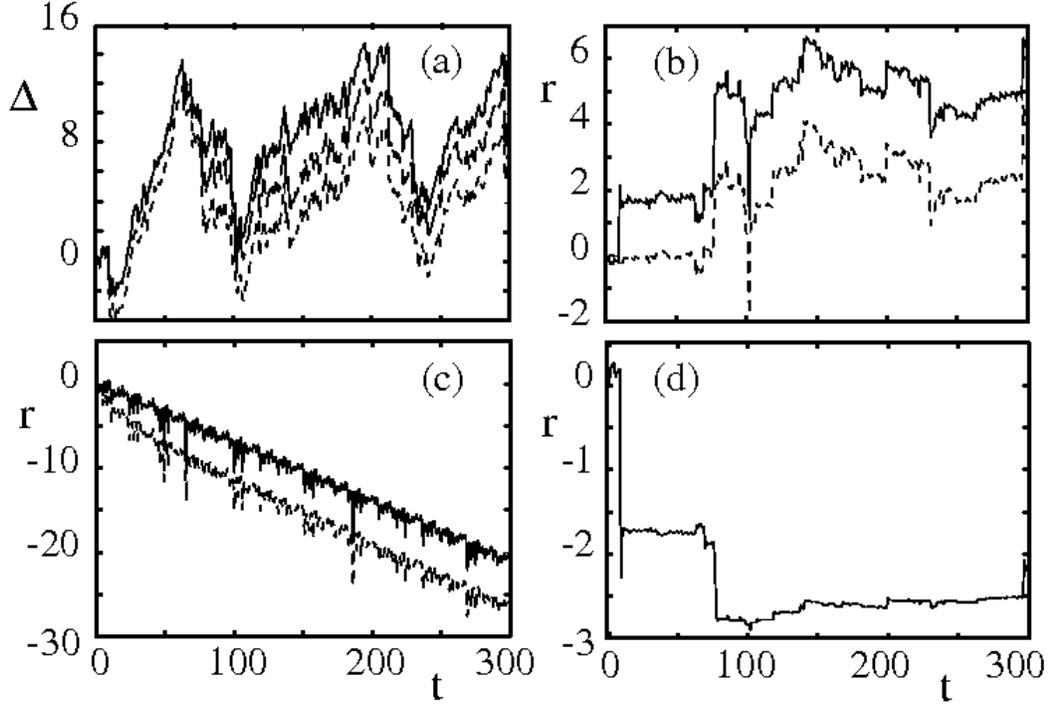} 
\end{center} 
\caption{Comparison of the two approximations, Eqs.~(\ref{exacttrj0},\ref{exacttrj01}), for the trace of $M_t(q_0,p_0)$ with the exact value.  In panel (a), the three curves are for K=8 with the initial condition $(q_0,p_0)=(0.3,0.6)$, $\Delta= \ln |M_t(q_0,p_0)|-t\ln(K/2)$, the exact trace is the solid line, the $j=0$ approximation is the dotted line (lowest), and the $j=0+1$ approximation is the dashed line.  In panel (b), $r= \ln\left[ |M_t(q_0,p_0)|/|\tilde M_t(q_0,p_0)|\right]$.  For the solid line, $\tilde M_t(q_0,p_0)$ is evaluated to the $j=0$ approximation, and for the dashed line, $\tilde M_t(q_0,p_0)$ is evaluated to the $j=0+1$ approximation.  Similarly to panel (b), in panel (d), the ratio of the two approximations is given.  Panel (c) is similar to panel (b), except that $K=9.28$ with the initial condition $(q_0,p_0)=(0.19,0.38)$ (which is a stable trajectory).  The solid line is for the $j=0$ approximation whereas the lower curve is for the $j=0+1$ approximation.}
\label{err1}
\end{figure}
Figure~(\ref{err1}) compares for a single trajectory the natural logarithm of the exact trace, the product form estimation of Eq.~(\ref{exacttrj0}), and the estimation of Eq.~(\ref{exacttrj01}) where the mean exponential increase has been subtracted from all three in the first panel.  In these panels several general features are readily seen.  As a function of time, the leading approximation follows the exact trace time step by time step except for sudden vertical displacements at time steps where the factor $|2-K\cos(2\pi q)|$ nearly vanishes.  In this example, the corrected approximation follows the exact trace longer in time before departing; although, this does not have to be the case.  A second point is that the exponential growth dwarfs the errors of the approximation.  Even though the approximation of Eq.~(\ref{exacttrj0}) is off by a factor of roughly $\sim{\rm e}^{4\leftrightarrow 6}$ before $t=300$, this is miniscule compared with the average exponential growth of the trace (in this example it is $\exp [t\ln K/2]= 4^t)$ and so it alters the stability exponent very little.  From the form of the leading correction, a Levy-flight nature might be expected for the corrections.  This can be seen in the natural logarithm of the ratio of Eqs.~(\ref{exacttrj0},\ref{exacttrj01}) respectively where only a couple of jumps dominate the differences between the zeroth and first approximation.  Although not shown, as K increases, a corresponding curve seems to be rather ``quiet'' punctuated by relatively fewer jumps.

These approximations naturally are worst for segments of trajectories that are stable or nearly stable.  For them $2-V^{\prime\prime}$ is repeatedly close to zero.  The large value of $K$ thus does not order corrections.  An example is given of a nearly stable trajectory in a region near a double saddle point bifurcation.  The value of $K$ is larger than used previously to illustrate the behavior of a highly unstable orbit.  Nevertheless, the approximation is far worse.  On the other hand however, it gives a stability (or Lyapunov) exponent  of about 0.067, a small number compared with $\mu_\infty \approx 1.53$.   As long as the percentage of stable and nearly-stable orbits is very small, the errors in using Eq.~(\ref{exacttrj0}) for all the orbits regardless of their nature are small after averaging over initial conditions. For the usual Lyapunov exponent there exists a ``local''  expression which is in some sense the equivalent of the $j=0$ expression for the trace we have just examined \cite{Fujisaka83}.

\subsection{The average finite-time stability exponents}

Much of what is known applies readily if a product form can be found for $\tr[M_t(q_0,p_0)]$.  This is precisely what the kicked rotor has for sufficiently strong potentials and times not too long.  There the $j=0$ term would be sufficient, and from Eqs.~(\ref{deflambda},\ref{exacttrj0}), the $\{\lambda_t(q_0,p_0)\}$ are approximately given by
\begin{equation}
\label{approx}
\lambda_t^0(q_0,p_0) = {1\over t} \sum_{i=0}^{t-1} \ln |2- V^{\prime\prime}(q_i)|
\end{equation}
The superscript $0$ indicates that it is the zeroth order approximation.  Assuming that the system is fully chaotic, the expectation value is given by integrating (averaging) over all initial conditions and  
\beq
 \langle \lambda_t^0\rangle = \int^1_0{\rm d}q \int^1_0{\rm d}p\ \lambda_t^0(q,p) = {1\over t} \sum_{i=0}^{t-1}\int^1_0{\rm d}q \  \ln |2- V^{\prime\prime}(q)|  = \int^1_0{\rm d}q \  \ln |2- V^{\prime\prime}(q)| 
\eeq
For the standard map, the result is $ \langle \lambda^0\rangle =   \ln {K\over 2}$; the integral is evaluated {\em exactly}.  Not surprisingly, this coincides exactly with the Lyapunov exponent approximation made by Chirikov \cite{Chirikov79} even though this result is for the stability exponent.  The phase-space mean of the finite-time stability exponent is {\em independent} of integration time and time is dropped from its notation.  This indicates that the mean is to within approximations made so far equal to $\mu_\infty$.  In contrast, the entire phase space average of the FTLE is not equal to its infinite time ergodic average, but has a transient that decays as a power law in time and in the absence of significant islands of regular motion goes as $1/t$.

\begin{figure}[t]
\begin{center} 
\leavevmode 
\epsfxsize = 15.0cm 
\epsfbox{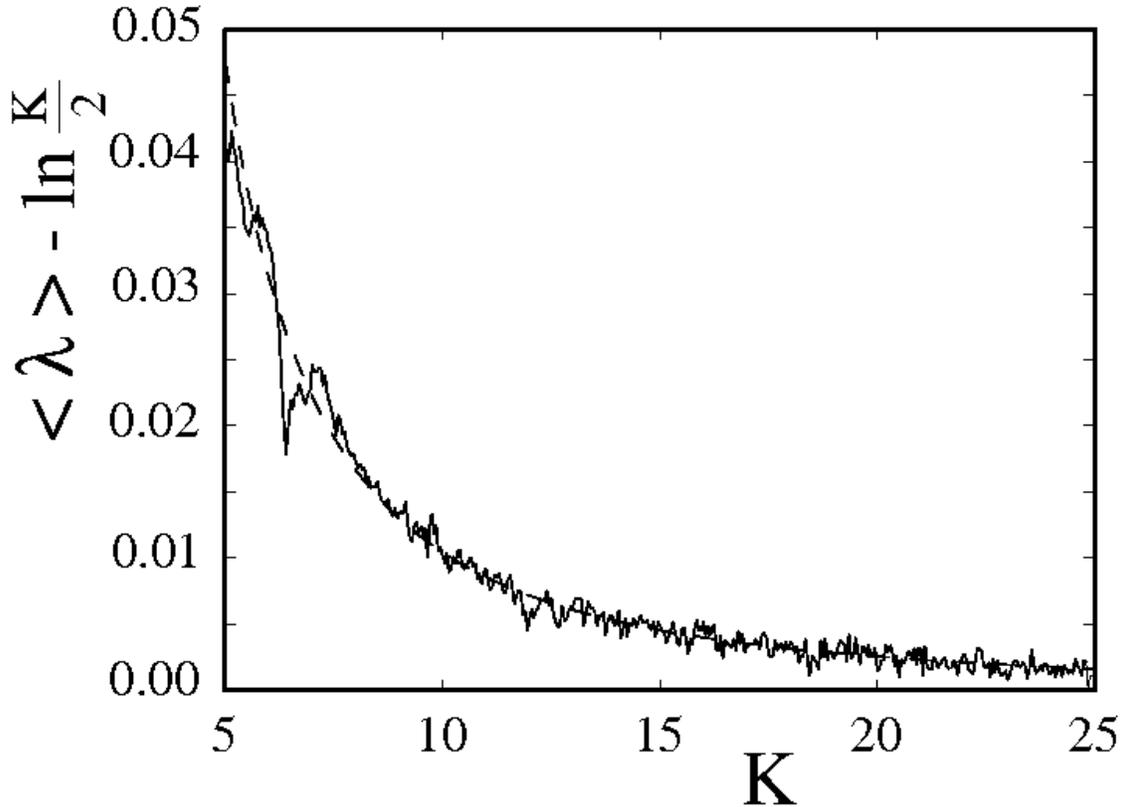} 
\end{center} 
\caption{The exact mean stability exponent calculated as a function of kicking strength with the zeroth order approximation, $\ln( K/2)$, subtracted.  Other than small oscillations, the difference is seen to be almost exactly equal to the first correction, $(K^2-4)^{-1}$, given in Eq.~(\ref{finally}).  This is plotted as the dashed curve .}
\label{cumul1}
\end{figure}

We plot the mean stability exponent in Fig.~(\ref{cumul1}).  
It is interesting that there is a simple systematic correction to the well-known formula of Chirikov above that is of the order of $1/K^2$, and that other than the value of $K$ (excluding sample size noise), there is essentially no indication of any other kind of dynamical information.

The origin of the corrections lie in the terms with $j\neq 0$ in the approximations to the exact trace, however its rigorous derivation seems fraught with difficulties.   A somewhat heuristic derivation of the $1/K^2$-correction to Chirikov's expression can be derived from Eq.~(\ref{exacttrj01}).  The product term gives an independent contribution and is exactly $\ln K/2$ as already noted.   Subtracting that term, one is left with the ergodic average of at least the $j=1$ term to determine the correction;
\begin{equation}
\delta \lambda (K) = {1\over t}\left<\ln \left|1 - \sum_i^{cyclic} {1\over  \left(2- K\cos 2\pi q_i\right) \left(2- K\cos 2\pi q_{i+1}\right) }  \right|\right> 
\end{equation}
We are not aware of a way to evaluate this integral without making a few simplifications.  Consider first that the correction is essentially being expanded about $K= \infty$ or in other words the expansion is about $\epsilon \equiv 1/K=0$.  In this limit at some fixed value $t$, the probability is vanishing that more than one factor of $|2- K\cos 2\pi q_i|$ for a given trajectory is close enough to zero that it can meaningfully contribute.  Let $q_0$ denote the $q$ value for the minimum of the set of $|2- K\cos 2\pi q_i |$.  Although that involves two terms in  the series, only one will dominate.  However, since there are $t$-terms in the cyclic sum, there are $t$ ways this can happen; together this is consistent with what happens in the subject of extreme statistics.  The basic expression to be evaluated is
\begin{equation}
\delta \lambda (K) \approx  \left<\ln \left|1 -  {1\over   \left(2- K\cos 2\pi q_0\right) \left(2- K\cos 2\pi q_1\right)}  \right|\right>  
\end{equation}
Due to the momentum not entering the expression, $\{q_0,q_1\}$ can be integrated as two independent variables each uniformly distributed on the interval $[0,1)$.  The next step follows from the discontinuity in the expression for the integral
\begin{equation}
\int_0^1{\rm d}q \ln \left| a\pm b\cos 2\pi q \right| = \left\{\begin{array}{lr}
\ln {|b| \over 2} & \qquad |b|\ge |a| \\
 \ln {|a| + \sqrt{a^2-b^2}\over 2} & |a|\ge |b| \\
\end{array}  \right.
\end{equation}
It is straightforward to verify that to the extent that the upper form of the integral applies, the expression for $\delta \lambda (K)$ vanishes.  Therefore it turns out that only values of $q_0$ for which 
\begin{equation}
-{1\over K-2} \le  2- K\cos 2\pi q_0  \le {1\over K+2}
\end{equation}
contribute a correction to the average stability exponent and the limits of integration on $q_0$ are restricted to that domain (actually there are two such domains).  Switching variables to $y=2-K \cos 2\pi q_0$, one has (accounting for both domains)
\begin{equation}
\delta\lambda(K) = {1\over \pi} \int_{-{1\over K-2}}^ {{1\over K+2}} {{\rm d}y \over \sqrt{K^2-(2-y)^2}}\left(\ln \left| {1-2y + \sqrt{(1-2y)^2-K^2y^2} \over 2}\right| - \ln {K|y|\over 2} \right)
\end{equation}
If the leading square root factor is expanded in a series of powers of $y$, the integrals can be analytically evaluated.  This gives,
\begin{equation}
\label{finally}
\delta\lambda(K) = {1\over K^2-4} + {19\over 6 (K^2-4)^3} + {4\over (K^2-4)^4} + {24 \over (K^2-4)^5} + \cdots
\end{equation}
As seen in Fig.~(\ref{cumul1}), the first term accounts completely for the overall correction to the Chirikov result to the accuracy level of the numerical calculations.   The $j=1$ trace correction appears to have given properly the stability (Lyapunov) exponent to $O(K^{-6})$.  This is interesting because it seems that it is not necessary to consider the $j=2$ and higher corrections to the trace whereas one might have naively expected the $j=2$ trace correction to give an $O(K^{-4})$ contribution.  Chirikov's accurate estimate for the Lyapunov exponent also has the same corrections derived above. This may be expected as at infinite times the Lyapunov and stability exponents indeed converge for the standard map. However, numerical calculations of the FTLE, even averaged over the entire phase space, mask this with transients that are of the order of $1/(t K)$. 

\section{Local properties of finite-time stability exponents}
\label{average}

The FTSE $\lambda_t(q_0,p_0)$ is also a {\em local} stability exponent as emphasized by its dependence on initial conditions.  The mean over a local ``area'' $\cal A$ that is much smaller than the  entire phase-space ($\Omega$)  includes transients that are power-laws in time.  Its interesting variations over phase space can be displayed by assuming the following particular form:
\beq
\label{ftav}
\br \lambda_t(q_0,p_0)\kt_{\cal A}= \dfrac{C_{\cal A}}{t^\gamma} + \ln(K/2) + {1\over K^2-4} +{\mathcal O}(K^{-6}).
\eeq
The area ${\cal A}$ is ideally a very small local region of the available phase space centered at $(q_0,p_0)$ and the final two terms are responsible for removing $\mu_\infty$, thus leaving just the fluctuating part.   Here $K>>5$ (in practice $K>6$) and the time $t$ is larger than the logarithmically short mixing time scale, $\tau_m= \ln (\Omega/{\cal A})/\mu_\infty\approx \ln(1/{\cal A})/\ln(K/2)$.  This time scale arises as the ``log-time'' during which the initial region of volume ${\cal A}$ through propagation essentially elongates to the extent that it enters each distinct local region of area ${\cal A}$ once on average.  The averaging in the LHS of the above expression is over the uniform Lebesgue measure, which is the invariant measure of the map.  Note that when the {\em entire} phase-space, i.e. the unit square for the kicked rotor,  is used for averaging the FTSE, the transients in time seem to be exponentially small and essentially $C_{\Omega}=0$. 

Expressing the finite time stability exponents in this way allows us also to draw attention to 
anomalous features that may be present for various values of $K$, especially those with small, but significant, stable islands in phase space that survive for large $K$, such as the accelerator modes. The quantity $C_{\cal A}$ fluctuates both in time and phase space point around which the averaging
is done, it can also have a strong dependency on $K$.  The index $\gamma$ serves to remove the principal non-fluctuating time scaling part of the FTSE and determines the rate at which it approaches the ergodic average.  For example, in the case when there are no apparent small islands  $\gamma=1$ (see text ahead), while in the presence of stable regions  $0< \gamma<1$.   The area ${\cal A}$ serves as a region of coarse-graining and it must be emphasized that unless we choose ${\cal A}$ to be the entire phase space ($\Omega$), the regions will exclude regular islands and trajectories. We present results for an exponent similar to $\gamma$ when studying the variance below. Previous studies of the FTLE have also presented such forms of convergence to the infinite time Lyapunov exponents \cite{Grassberger88,Abarbanel91, Goldhirsch87}, but typically in dissipative systems.

\begin{figure}
\includegraphics[width=6.5in,height=4.0in]{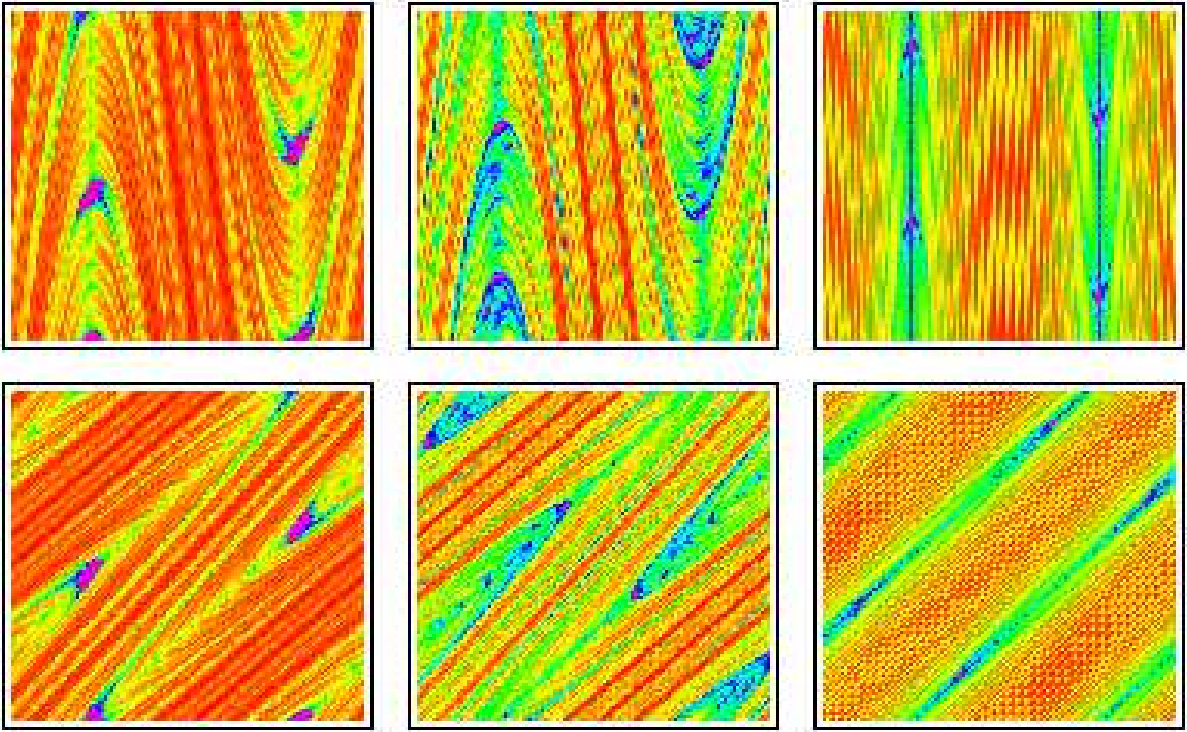}
\caption{A density plot of the  average $\br \lambda_{\pm 10}(q_0,p_0) \kt_{\cal A}$ (top, bottom respectively) as a function of $(q_0,p_0)$. High to low density is indicated as a color transition from purple to red. The area ${\cal A}$ over which the averages are calculated are squares of area $0.005^2$. The case $K=10$ (center column) has no apparent stable islands, whereas at $K=9.26$ (left column) there are two small islands each of period four  where the density is enhanced and purple (see the next figure for a magnification of the dynamical structure underlying the purple zone).  Note the prominent linear structures for a large value of $K\, ($=40$)$ in the right column.}
\label{phsp}
\end{figure}

To get an idea of the complexity of the phase space distribution of the FTSE as we move $(q_0,p_0)$ around, always averaging over a small area ${\cal A}$ of phase space centered at this point, we display density plots in Fig.~(\ref{phsp}).  
The figures show forward and backward in time evolution for three cases of $K$, namely $9.26$, $10$, and $40$ respectively.  The intricate structures that arise highlight areas of reduced hyperbolicity and hence smaller FTSE.  For the case of forward evolution in time, trajectories which are trapped in more stable areas or propagate into those areas have lowest stability and appear purple.  Those trajectories propagating into more stable zones must do so in a direction aligned along the greatest exponential compression.  This direction is determined by the stable manifolds of the short periodic orbits.  Thus, one sees a structure that mirrors the foliation of phase space by these stable manifolds in the top row of the figure.  Similarly, for reverse evolution in time, one sees a structure that mirrors the corresponding unstable manifolds in the lower row.  In the first column, the phase space has small islands, while the center column, $K=10$, does not have such prominent islands. The absence of a detectable island of regular motion results in a less contrasted variation of the FTSE.  At $K=40$ (right column), which is very strongly chaotic, there are linear regions of low FTSE that are clearly visible.  However,  any other regions have more or less merged into the background. These linear regions coincide with the $K$ independent  borders of the non-hyperbolic regions for a one-step forward (the vertical lines at $q=1/4$ and $q=3/4$) and one-step backward iteration (the lines $q-p-1/4=0 \,\mbox{mod}\, 1$ and $q-p-3/4=0 \,\mbox{mod}\, 1$).  The fact that these lines are not dense in the phase space and survive for long times at large $K$ underlies the coexistence of strongly chaotic dynamics with the fact that the map may not possess a single $K$ value for which it becomes mathematically fully hyperbolic.  Similar plots for the FTLE have been plotted earlier in the contexts of area-preserving maps for smaller values of $K$ \cite{Eckhardt93} and passive scalar advection \cite{Lapeyre02}. 
 
\begin{figure}
\includegraphics[width=6.0in,height=5.0in]{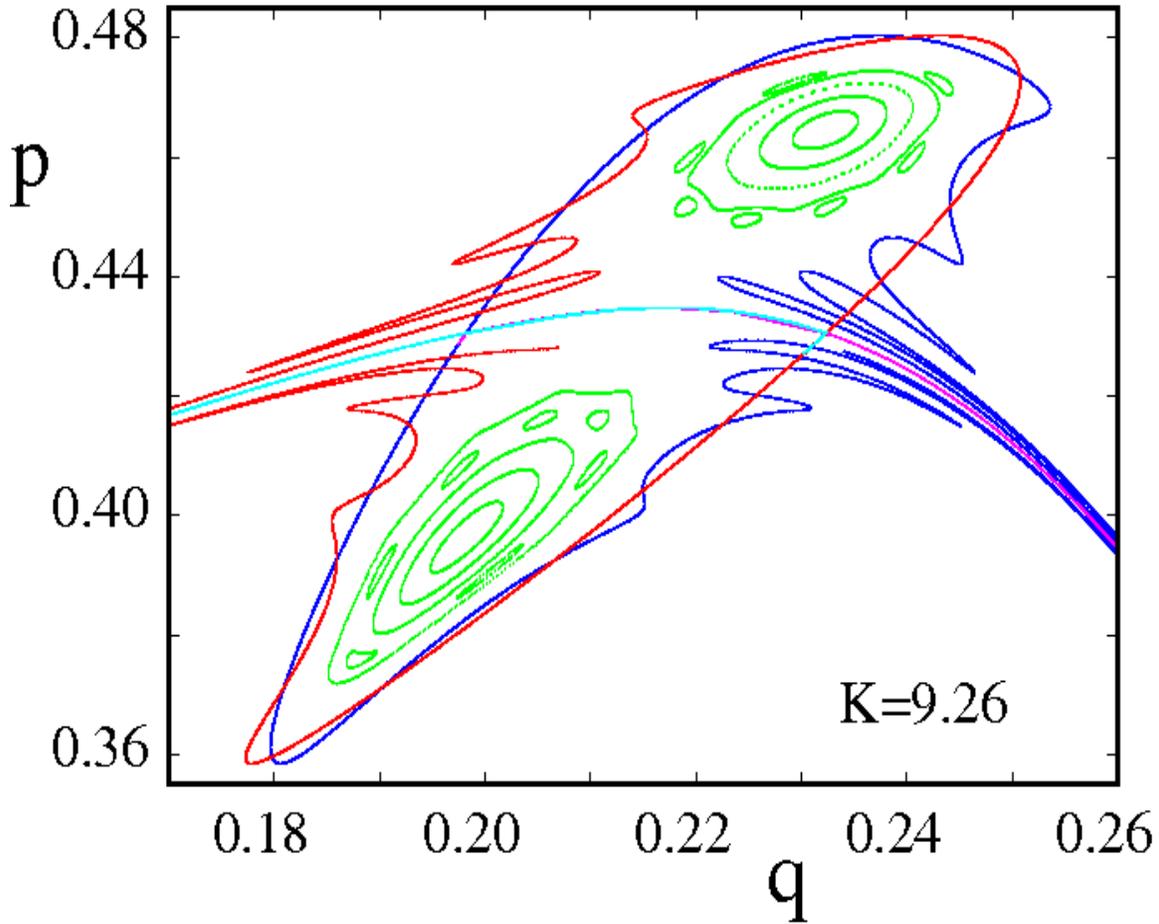}
\caption{Expanded plot of two islands of period-four regular motion and the stable and unstable manifolds associated with two unstable orbits born at the same double saddle node bifurcation (K=9.20667769750) as the islands themselves.  Only one of this structure's four images is plotted.  Transport in and out of the inner region is severely limited by the small turnstile fluxes visible.  }
\label{island926}
\end{figure}

It is possible to use this plot to help locate small islands.  Investigating the purple zones seen in the left column of Fig.~(\ref{phsp}) more closely reveals the interesting structure found in Fig.~(\ref{island926}).
It arises for $K$ values greater than approximately $9.2$ at which point there is a double saddle node bifurcation.  As $K$ increases from the bifurcation, the structure grows to a maximum size near $K=9.26$.  The full structure includes not only the small stable islands, but also the short unstable periodic orbits along with their homoclinic tangles, i.e. crossings of stable and unstable manifolds.  The turnstiles formed by these crossings limit the rate trajectories can move in and out of the region surrounding the islands of stable motion \cite{MacKay87}.  Therefore in this structure, it is not just the immediate interface between the chaotic and regular motion which is generating trapping, although that, of course, is happening additionally just beyond the outermost tori pictured.  In fact, chaotic trajectories may remain trapped inside the tangle for hundreds or more time steps in the roughly 2/3 portion outside the islands themselves.  For this reason, the entire structure was visible in Fig.~(\ref{phsp}), not just the islands, and multiple trapping time scales would be expected.  One of the more unusual features is a highly asymmetric instability (factor of 1000 or so) in two of the stable and unstable manifolds pictured.  The interior middle line is actually a stable and unstable manifold whose turnstile is so tiny as to appear as a single line dividing the interior, thus making transport between the upper and lower region nearly non-existent.  

\section{Fluctuations of the finite-time stability exponent}
\label{fluc}

The statistical properties of $\{\lambda_t(q_0,p_0)\}$ can be studied via the construction of a  probability density $\rho_{\lambda_t}(x;t)$, i.e. the probability of finding $\lambda_t$ in the interval ${\rm d}x$ centered at $x$.  For a large class of chaotic systems, a great deal is known about general properties of $\rho_{\lambda_t}(x;t)$, although strictly speaking just for the FTLE.  The approach to a unique Lyapunov exponent is connected to  $\rho_{\lambda_t}(x;t)$ approaching a $\delta$-function limit as $t\rightarrow\infty$.  Through the study of the density's cumulants for fully hyperbolic systems, one sees that there is a slow approach to an ever narrower, roughly Gaussian density as $t\rightarrow\infty$ \cite{Ott97}.  A detailed study with analytic results has also been performed for ``white-noise'' random media \cite{Schomerus02}.  Interestingly enough, it is also known that this limiting process is too slow to understand the behavior of the generating function for the cumulants $\nu_n(t)$
\begin{equation}
\label{cumulantsgen}
\ln \left< \exp mt \lambda_t \right> = \sum_{n=1}^\infty \nu_n(t) {(mt)^n\over n!} \approx \ln \left< \left| \tr(M_t) \right|^m \right> 
\end{equation}
which is also a quantity of direct interest in semiclassical studies or the theory of Anderson localization.   This relation should be thought of as an expansion in the real variable $m$.

If one were to calculate the full probability density  $\rho_{\lambda_t}(x;t)$ for the kicked rotor, one would see that it qualitatively takes on one of  two forms.  If the kicking strength is set to a value at which there is no discernible evidence for any stable orbits (no detectable non-exponentially decaying dynamical correlations), the density has a well defined peak near the Lyapunov exponent, and is not too far off a Gaussian form.  If the kicking strength is set to a value at which both stable and strongly chaotic trajectories co-exist in the phase space, then it takes on a bi-modal form with a peak near the mean Lyapunov or stability exponent and another representing algebraic behavior near zero \cite{Sepulveda89}.  The density is continuous between the two peaks.  Nevertheless, as $t\rightarrow \infty$, a unique Lyapunov exponent emerges albeit much more slowly.  This has a profound effect on the time scaling properties of the density or more specifically its cumulants (moments) with the exception of its mean.  The algebraic peak and the `neck' connecting the two peaks disappears in the $t\rightarrow \infty$ limit very differently than the rate at which the width of the main peak narrows.  Time scaling therefore gives a great deal of information about non-hyperbolicities.

\subsection{Time scaling of the variance}

We proceed next to a more detailed analysis of the variance (second cumulant) by first deriving the contributions that neglect dynamical correlations, which leads to an explicit formula for the standard map; the first cumulant (the mean) is already treated above.  It is shown  that while the general time scaling of the variance is $1/t$, this is significantly modified by the presence of even very small stable islands, thus providing a practical way of detecting a `barely' mixed phase space.  We anticipate that this would be especially useful in its higher-dimensional generalizations \cite{Tomsovicinprep}.  A first study appears to support our expectations \cite{Beims07}.

The variance of the FTSE over $\Omega$ can be expressed simply using the approximation of Eq.~(\ref{exacttrj0}) for the exact trace; the exact one, $\nu_2(t)$, differs from its approximation above, $\nu^0_2(t)$, by very small amounts which  can be ignored.  One finds:
\beq
\label{vardefn}
\nu^0_2(t)=\int_{0}^{1}\int_{0}^1 dq_0 dp_0 \, \left( \df{1}{t} \sum_{i=0}^{t-1} \ln|2-V''(q_i)|\right)^2 -(\nu_1^0)^2
\eeq
The variance may be be separated into a term that is explicitly independent of dynamical correlations and a series of terms that contain the contributions of dynamical correlations, which respectively are:
\begin{eqnarray}
\nu_2^0(t)&=&  {1\over t}  \int_0^1\int_0^1\ {\rm d}p {\rm d}q \left(\ln^2\left| 2- V^{\prime\prime}(q)\right|   -(\nu_1^0)^2\right) \nonumber \\
&& \  + {2\over t^2} \sum_{\tau=1}^{t-1}(t-\tau)  \int_0^1\int_0^1\ {\rm d}p_0 {\rm d}q_0 \left(\ln\left| 2- V^{\prime\prime}(q_0)\right|  \ln\left| 2- V^{\prime\prime}(q_\tau)\right| - (\nu^0_1)^2\right)
\end{eqnarray}
The contribution of the first term above is
\begin{equation}
\label{pi212}
\dfrac{1}{t}{\cal X}_2 \approx  \dfrac{1}{t}\int_0^1 {\rm d}q (\ln|K\cos(2\pi q)|-\ln(K/ 2))^2  = {\pi^2 \over 12 \, t} = \dfrac{1}{t} 0.822\ldots.
\end{equation}
Here we have introduced ${\cal  X}_2$  as a term  that does not carry correlation information.  

\begin{figure}[t]
\begin{center} 
\leavevmode 
\epsfxsize = 15.0cm 
\epsfbox{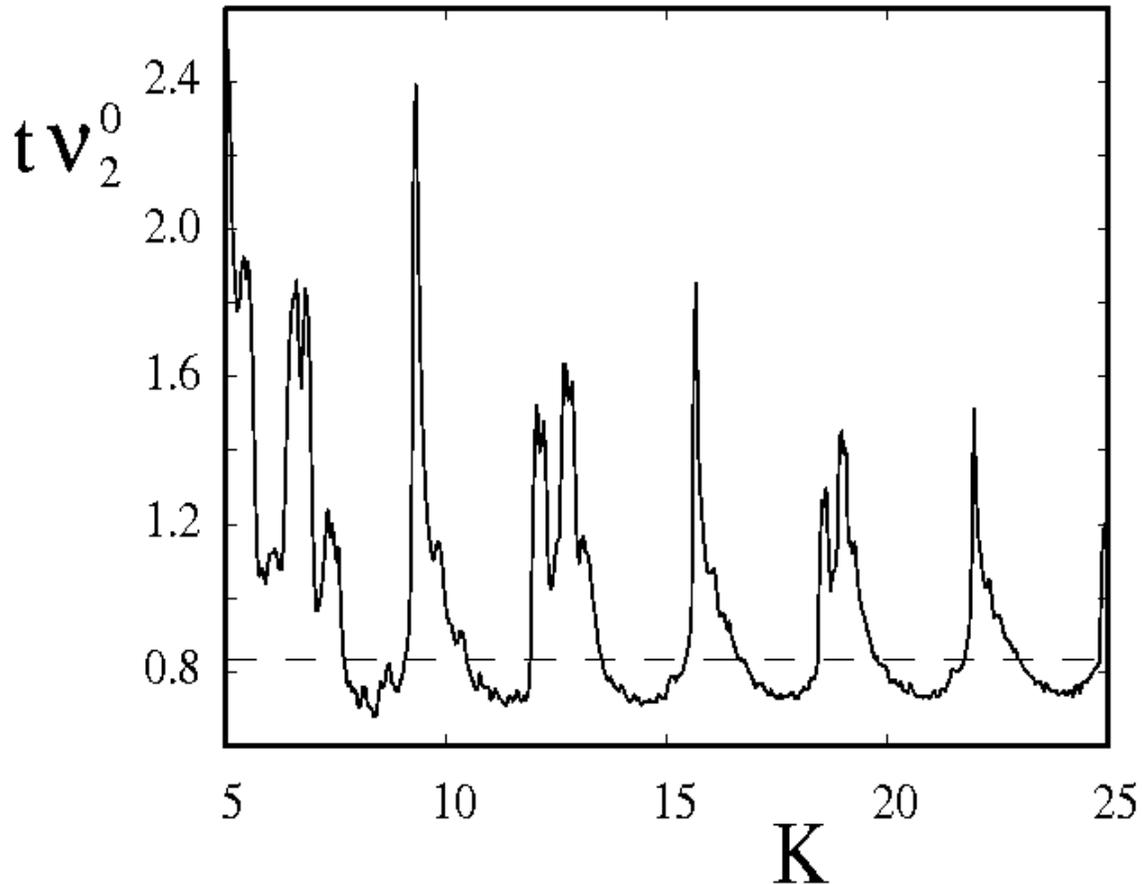} 
\end{center} 
\caption{The $j=0$ order calculation of the variance as a function of kicking strength.  It is multiplied by $t$ and averaged over $t=150$.  It illustrates that the variance is very sensitive to dynamical correlations.  The dashed line at $\pi^2/12$ is the value of the integral from Eq.~(\ref{pi212}). }
\label{cumul2}
\end{figure}

Depending on the system, strong behavior connected to dynamical correlations but not reflected in the mean, is possible in the variance, which the summation above makes clear.  Note that the dynamical correlation function is bounded for a broad class of potentials.  Furthermore, as long as the correlations decay faster than $\tau^{-1}$, they cannot alter the basic time-dependence of the variance.  If they decay more slowly, they could well become the leading contribution in time.  As seen in Fig.~(\ref{cumul2}) the variance is strongly dependent on the value of the kicking strength $K$ and shows marked peaks at various values. 
These include conventional accelerator modes as well as other stable islands, such as those illustrated 
in Fig.(\ref{island926}) that appear near $K=9.3$. 

\subsection{The variance as a sensor for small elliptic islands}
\label{variance}

As just noted, the variance scales with time differently from the $1/t$ law above as  a result of  correlations decaying as slow or more slowly than $\tau^{-1}$.  For the kicked rotor, consider ensembles of trajectories that originate from small regions $\cal A$ of the phase space rather than the whole space; the reduced averaging region is motivated by the interest in a process that more practically generalizes to systems with greater numbers of degrees of freedom.  The average behavior $\br \lambda^0_t(q_0,p_0)\kt_{\cal A}$ may obscure the scaling by the presence of the infinite-time asymptotic value, whereas the fluctuations die out asymptotically and are more stable for evaluating power laws.  Although, note that ``wiggles'' in the FTLE were attributed to stability islands in an earlier work on kicked tops \cite{Constantoudis97}. The variance can be written
\beq
\label{varpower}
\mbox{Var}[\lambda^0_t(q_0,p_0)]_{\cal A} =D_{\cal A}/t^\alpha.
\eeq
where $\nu^0_2(t)=\mbox{Var}[\lambda^0_t(q_0,p_0)]_\Omega$.  After accounting properly for the $t^{-\alpha}$ dependence, one sees that $D_{\cal A}$ fluctuates as a function of time similarly to a diffusive variable. Generally speaking, the greater the time range over which $\alpha$ is calculated and the greater the number of trajectories run, the greater precision with which $\alpha$ is determined.  In this regard, the simplicity of the $j=0$ approximation is quite helpful because it makes it possible to increase both in a practical sense.  One suspects that the smaller and less influential the island of regular motion, the closer $\alpha$ approaches unity and thus $\alpha$-accuracy is directly linked to how small an island one can detect in this way.   Earlier works such as  \cite{Grassberger85,Horita90} have studied scaling laws that arise in the fluctuations of the FTLE and have already emphasized the role of sticky regions and islands.  Ahead, we show islands of measure roughly $0.01\%$ of $\Omega$  that were found with the FTSE time scalings relatively easily.

\begin{figure}
\includegraphics[width=6.0in]{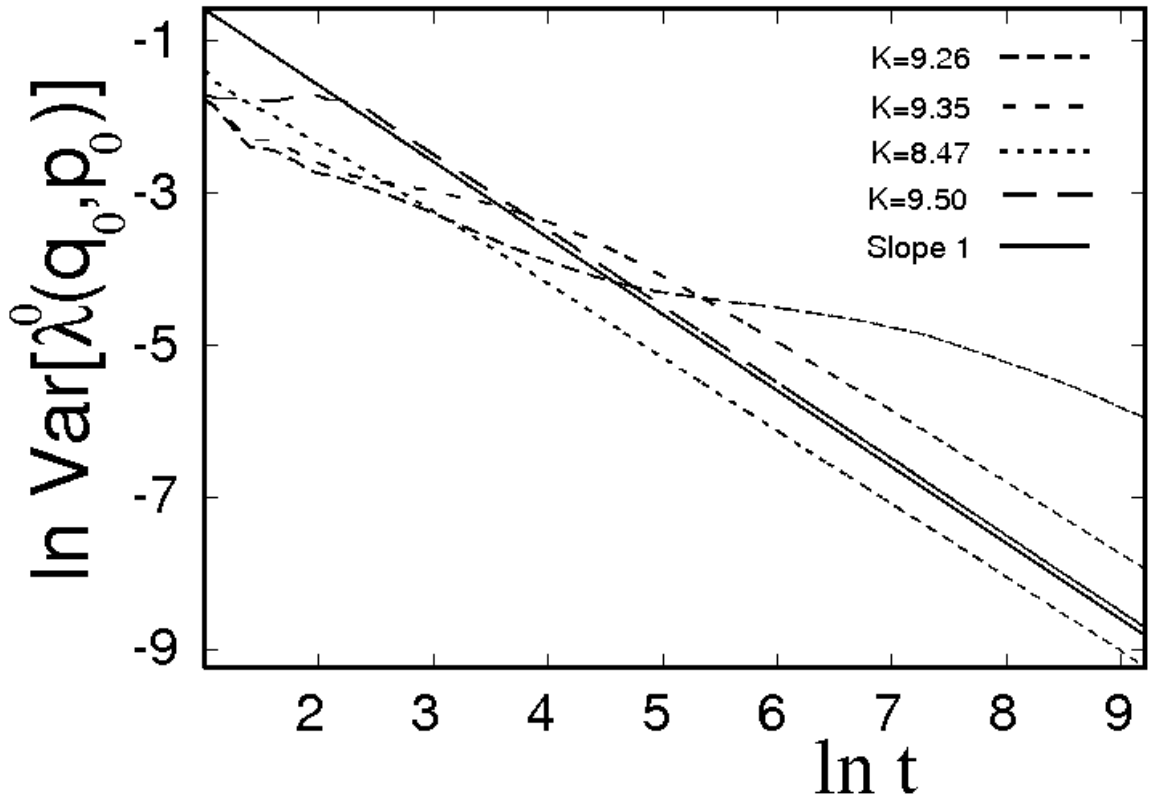}
\caption{The scaling of the variance for an ensemble of $40000$ initial conditions uniformly distributed in the square ${\cal A}=\{(0.10,0.10),(0.15,0.15)\}$. }
\label{varscal}
\end{figure}

\begin{figure}
\includegraphics[width=6.0in]{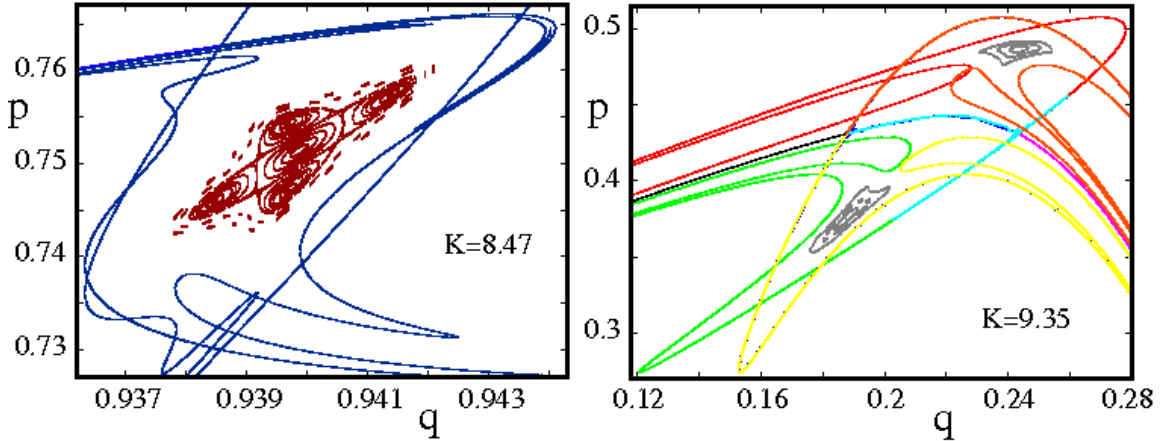}
\caption{Small regular islands found at $K=(8.47, 9.35)$ and their surrounding homoclinic tangles. Not only are the islands much smaller than for $K=9.26$, but the turnstile fluxes are larger relative to the measure of the enclosed area, making the flow much less restricted in the neighborhoods of the islands.  The measure of the regular region for $K=8.47$ is roughly $0.01\%$ of $\Omega$ counting all the unique iterations of the structure.}
\label{island847}
\end{figure}

In Fig.~(\ref{varscal}) we show the scaling of the FTSE variance for a few values of the kicking parameter compared with $\alpha=1$.  
The regular island structures of Fig.(\ref{island926}) at $K=9.26$ have a very significant effect on the time scaling.  One also sees the appearance of more than one time scale, i.e. slope, depending on the time regime.  The turnstiles previously pictured would at first lead to only small numbers of trajectories entering into the interior region of the homoclinic tangle, and one would expect $\alpha$ to begin close to unity.  Later, a certain proportion would get trapped inside for up to hundreds or thousands of time steps.  During this period, the greatest proportion would resemble nearly stable orbits and $\alpha$ would deviate most from unity.  Further in time, only the trapping right at the regular island-chaos interface would contribute to deviations from unity and the slope increases somewhat again.   It is known that different trapping mechanisms can co-exist in  phase space and lead to a ``multifractal'' process \cite{Zaslavsky02}.  This particular example is instructive because the turnstile time scale is fairly well separated from the others.  The smaller the stable islands, the more the power laws recover the uncorrelated $1/t$ form.   At $K=(8.47, 9.35)$  there are only small detectable deviations from unity, and in both cases it is possible to locate very small islands; see Fig.~(\ref{island847}).
At $K=9.5$ we could not locate islands in this way and the exponent $\alpha$ is as close to unity as the precision of the calculation.

It is worthwhile remarking that while accelerator modes give rise to anomalous diffusion (in momentum) in the standard map on the cylinder \cite{Ishizaki91}, the different scaling regimes discussed above hold whether the stable regions are accelerator modes or not.  However, the suppression of the rate of the variance's approach to zero and the average Lyapunov or stability exponent to the ergodic average in the presence of small islands has similar origins.  Namely, they are due to trajectories spending intermittently long intervals of time near sticky islands or regions of almost marginal stability and nonhyperbolicity.  

We point out that although power-laws such as that given in Eq.~(\ref{varpower}) are susceptible to
variations in the initial position $(q_0,p_0)$, to changes of the trajectory sampling, and to the time window in which they are under study, the $\alpha$ deviations from unity are robust structures and can be correlated to phase space structures.  
\begin{figure}
\includegraphics[width=6.0in]{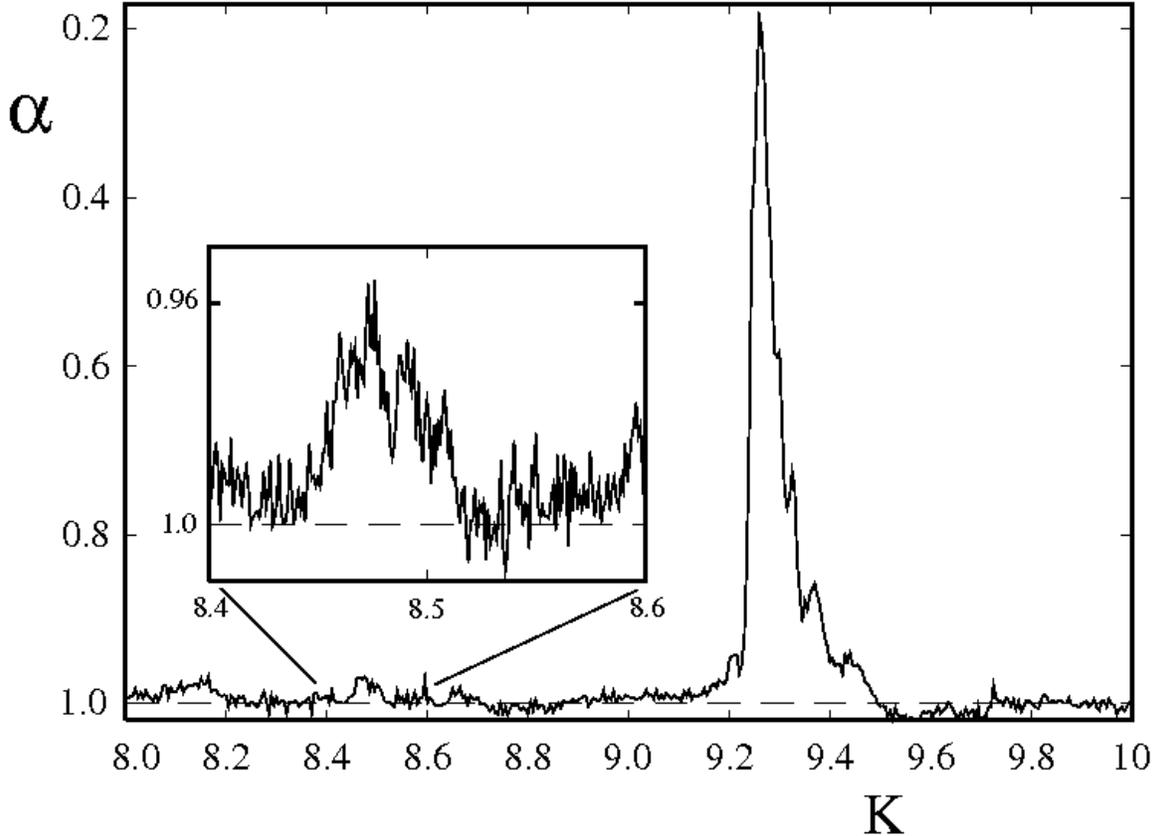}
\caption{The exponent $\alpha$ in the decay of the variance of the FTSE as a function of the kicking strength $K$. The ensemble is the same as that used in Fig.~(\ref{varscal}).  A region around $K=8.47$ is shown at a resolution of $0.001$ in $K$.}
\label{exponent}
\end{figure}  
In Fig.~(\ref{exponent}) we show the variation of $\alpha$ as  $K$ is tuned through 
a range which has small stable regions.  The ``normal '' exponent $\alpha=1$ is 
realized to a very good approximation for most values of $K$ indicating the general absence of
 significantly large, low instability traps. The prominent deviation around $K=9.3$ is due to the structure of Fig.~(\ref{island926}) born out of a double saddle node bifurcation around $K=9.2$ and it persists to near $K=9.45$.  It grows to its largest size around $K=9.26$ where the variance decays the most slowly, $\alpha\approx 0.16$.  In terms of the phase space area, the islands occupy approximately $0.5\%$ of $\Omega$ at $K=9.26$ and  $0.2\%$ at $K=9.35$ (for which $\alpha\approx 0.88$).   Notice robust, though small, deviations from unity at several other values of $K$.   For example, even for an island as small as that found at $K=8.47$ ($0.01\%,\ \alpha\approx 0.96)$, there is a range of $K$ values where $\alpha$ is significantly and consistently different from unity; the inset shows a finer scan of this region.  The fine oscillatory structure that  is apparent here could be reflecting the fractal behavior of such exponents themselves which in turn are due to bifurcations in the map that produce or destroy very tiny islands almost continuously.  The Newhouse phenomenon wherein homoclinic tangencies produce an infinite number of sinks is replaced in conservative scenarios by a large number of elliptic islands \cite{Duarte94}.  Knowing that there is possibly an island somewhere in phase space does not locate it, and a blind search could well be futile.  Some direction is important.  The islands are found more easily by searching around regions of the lowest stability exponent, regions that are highlighted in figures such as Fig.~(\ref{phsp}), and correspond to zones of low hyperbolicity.  Note one last observation, for the three islands constructed, $K=(9.26,9.35,8.47)$, the degree of $\alpha$'s deviation from unity and the measure of the islands are in corresponding order, smallest to largest.

\subsection{The higher order cumulants}
\label{higherorder}

Similar more complicated dynamical correlation expressions enter for all the higher cumulants.  They naturally weight more the extremes of the probability density, and thus, one might expect them to be even more sensitive to stable islands.  To be useful, they also need a standard time scaling for their correlation independent components as found for the variance.  In fact they do have a simple time dependence and it involves generalizations of ${\cal X}_2$. By defining:
\begin{equation}
{\cal X}_n= \int^1_0\int^1_0 {\rm d}p {\rm d}q\  \left(\ln \left| 2- V^{\prime\prime}(q)\right| -\nu^0_1\right)^n 
\end{equation}
the correlation-independent component of the higher order cumulants are expressible as polynomials in these quantities with a single overall power-law decay in time and no other parametric dependencies.  Details are presented in Appendix B, but note here an interesting consequence of these expressions that imply a special feature found in any of the 2D kicked rotors.  Substitution gives
\begin{equation}
\label{calx}
{\cal X}_n  \approx \int_0^1 {\rm d}q (\ln|K\cos(2\pi q)|-\ln{K\over 2})^n =  \int_0^1 {\rm d}q (\ln|\cos(2\pi q)|+\ln 2)^n
\end{equation}
which shows that to the extent that dynamical correlations can be ignored, every cumulant is independent of the overall kicking strength.  The probability density thus has a single functional form for all kicking strengths.  Therefore, although the Lyapunov exponent grows logarithmically with $K$, the probability density of FTSE retains the same shape and width (it just translates with $\nu_1$ as $K$ increases).  Assuming the kicking strength is large enough that the system is strongly chaotic, any $K$ dependence of the probability density, other than where it is centered, is due to variation of dynamical correlations and oscillatory.  

\begin{figure}[t]
\begin{center} 
\leavevmode 
\epsfxsize = 15.0cm 
\epsfbox{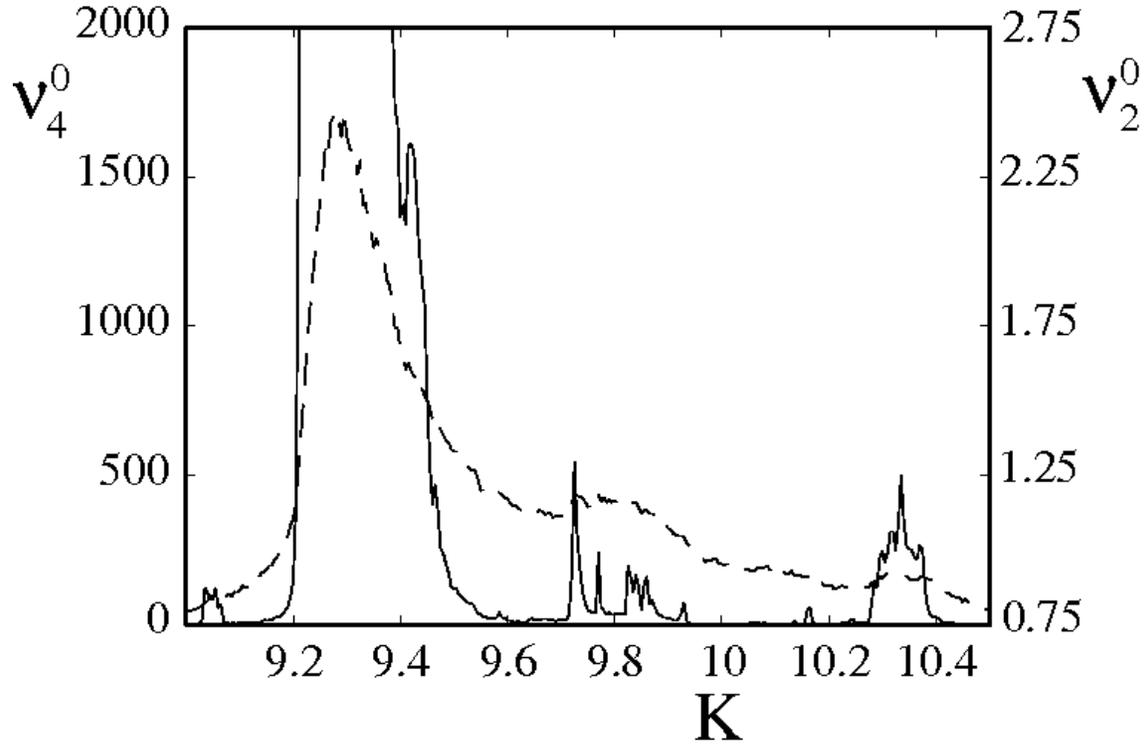} 
\end{center} 
\caption{The fourth and second cumulants as a function of $K$ using the $j=0$ approximation.
The solid line is the fourth cumulant and the dashed line is the second.  }
\label{cumul4}
\end{figure}

Even stronger dynamical correlations and more sharply defined fluctuations show up in the higher  cumulants.  In Fig.~(\ref{cumul4}), the fourth and second cumulants are compared.
The fourth cumulant displays much sharper features than the second.   On the other hand, the features are present in the same regions of $K$, so the connection to stable islands is the same as for the variance.  We did not carry out the corresponding time scaling calculations as we have no reason to doubt that they reflect the stable structures for the same reasons as for the variance.

\subsection{The fluctuations in $1/|\mbox{Det}(M_t-I)|$}

\begin{figure}
\includegraphics[width=6.0in]{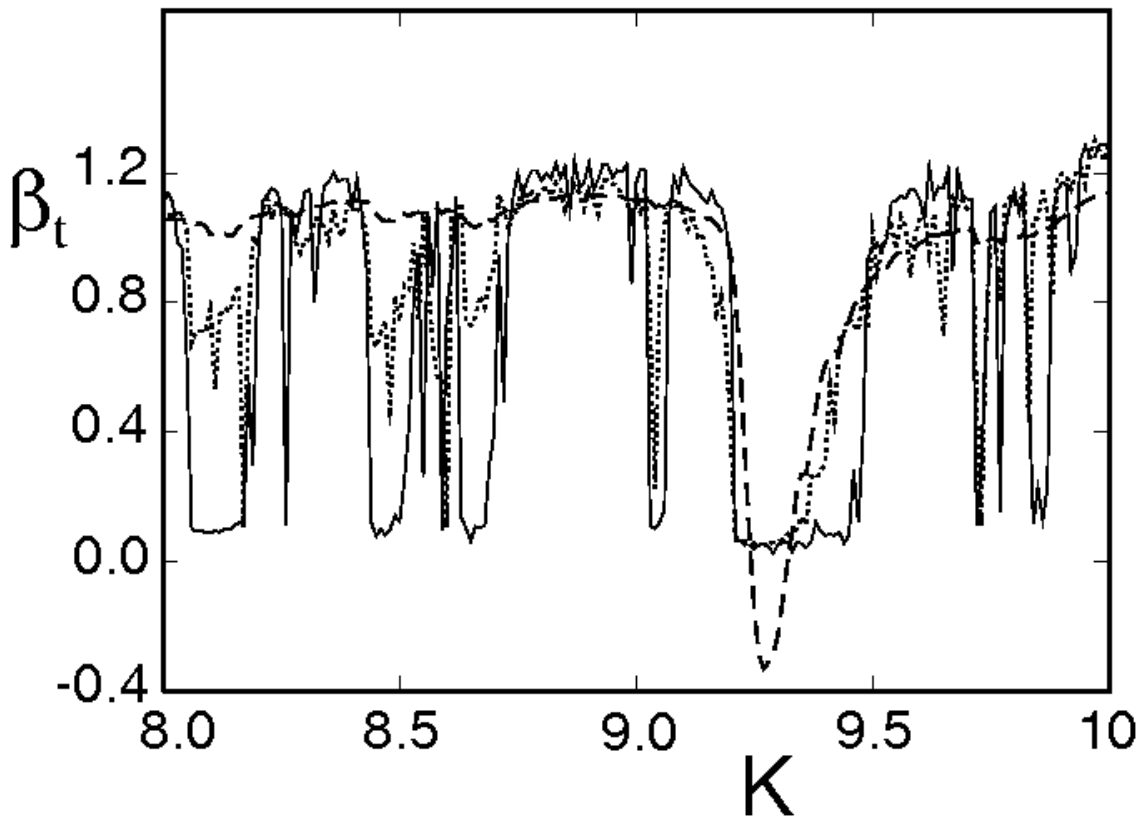}
\caption{The quantity $\beta_t$ as a function of $K$ at $t=100$. The averaging involved $40,000$ initial 
conditions spread uniformly across phase space.  The solid line is the exact trace.  The relatively smoothly behaving dashed line is what the result would be for a normal variable, i.e.~$\beta_t=\nu_1-t\nu_2/2$.  The dotted line derives from the $j=0$ approximation.  In principle, the approximation's errors are much worse for a quantity such as $\beta_t$ relative to one of the low order cumulants, $\nu_n$.}
\label{invtr}
\end{figure}

Since higher cumulants carry information and have increased sensitivity to stable structures, a quantity that contains information from all the cumulants has the potential to be roughly optimal for its sensitivity to stable structures.  Through relations such as Eq.~(\ref{cumulantsgen}), one sees that $1/|\mbox{Det}(M_t-I)|$ fits into this category.  Furthermore, it is a quantity of considerable interest in semiclassical theory, where the square root occurs as the amplitude along classical paths.  We therefore define 
\beq
\label{beta}
\beta_t=\dfrac{-1}{t} \ln \left \langle \dfrac{1}{|\mbox{Det}(M_t-I)|} \right \rangle 
\eeq
In a uniformly hyperbolic system, $\beta_\infty$ is the Lyapunov exponent.  However, if the hyperbolicity is not uniform, as is generically the case, then $\beta_t$ is less than the corresponding stability exponent.  In the presence of stable motion, it must tend toward vanishing.  The averaging above can be done over any ensemble of phase space trajectories over time $t$; we use a uniform density over $\Omega$. Shown in Fig.~(\ref{invtr}) is the variation of this quantity
across a range of $K$ values.  Its extreme sensitivity to small islands and sticky regions
of phase space is quite apparent. This may be expected as the denominator can come
arbitrarily close to zero in the neighborhood of points undergoing bifurcations. We show two 
calculations in the figure, one of which is an ``exact'' evaluation based on finding the trace 
of $M_t$ at large times, using a determinant method used initially for periodic orbits \cite{Bountis81, Greene79},  and the other is by using the product ($j=0$) approximation in Eq.(\ref{exacttrj0}). The latter procedure is a stringent test for this approximation and it is interesting that it picks out almost all of the ``hotspots''.  It is however much faster and easier to implement than the exact method. The smooth line shown is what one may expect from a normally distributed variable.

On comparison with $\alpha$ and with the variance itself, we find that the ranges of $K$ where $\beta_t$ deviates from that expected of a normal random variable, is also reflected in these quantities. Note the consistency in all these measures with respect to structures due to stable, if small, islands in the phase space.  For example, the structures around $K=8.47$ reflect those already illustrated in Fig.~(\ref{exponent}), while those around $K=8.66$ correspond to similar islands.  We again expect that higher dimensional generalizations of the approximation $j=0$ studied here will provide a method of ascertaining higher dimensional, hard to detect, `weakly' mixed phase spaces.

\section{Discussion}

The consideration of finite-time stability exponents has some advantages vis-a-vis finite-time Lyapunov exponents.  For one, they are more directly related to quantities that appear naturally in quantum mechanics.  For another, their generalization to many-degree-of-freedom systems tend to the Kolmogorov-Sinai entropy in the $ t \rightarrow \infty $ limit.  There is a local approximation just as there is for FTLE, but the time-dependent transient may disappear exponentially fast in many cases for FTSE and not FTLE, depending on averaging techniques employed.  Corrections to the local approximation are also relatively simple and the first one is sufficient to calculate improved estimates of Lyapunov exponents.  We showed that this gives a leading correction to $\ln {K\over 2}$ of exactly $(K^2-4)^{-1}$ for the standard map.  In fact, although not explicitly mentioned, we explored a wide variety of techniques all of which failed to give this correction properly (most vanished or diverged).  The corrections themselves are interesting as well.  One would guess that they should lead to a Levy-process and numerically this is borne out.  At least for low-order cumulants in the large kicking strength regime, the local approximation along with the first correction gives the full picture, and for many quantities the local approximation is more than sufficient.

For the standard map, the mean (first cumulant) FTSE converges rapidly to the Lyapunov exponent.  This is understood to be a consequence of $C_{\Omega}$ tending exponentially quickly to zero with time in Eq.~(\ref{ftav}).  The local approximation gives a mean FTSE that is independent of direct dynamical correlation contributions.   Indirectly, in the mostly chaotic regime, the exclusion of chaotic trajectories from small regular regions (islands) would slightly alter the ergodic averaging, but never more than the relative measure of those regions.  Consistent with this picture, other than small deviations, one sees very little effect of small islands on the mean FTSE.  

The fluctuations behave quite the opposite.  They can be investigated through the creation of a probability density for the FTSE.  In the absence of small regular regions, the probability density takes on a single form, independent of system parameters (such as the kicking strength), which slowly shifts towards a Gaussian form as $t\rightarrow \infty$ of width $\pi^2/12$.  Therefore, all deviations from this form must be due to dynamical correlations, and in particular, correlations introduced by the existence of small stable islands of motion.  Since very strong $K$ dependence is seen, the higher cumulants are very sensitive sensors of the presence of such non-hyperbolicity in the system.  As such, they can be used as `detectors'.  An important quantity which shows up in semiclassical theories, ${\rm Det}|M_t-1|^{-1}$, in principle, contains information carried in all of the cumulants.  In fact, it may be the most sensitive detector of dynamical correlations induced by tiny regions of non-hyperbolicity in the dynamics.  All of the techniques and main results discussed have their generalizations to many degree-of-freedom systems.  Those studies are currently underway \cite{Tomsovicinprep}.

\vskip .4cm
{\bf Appendix A}
\vskip .4cm

  The relation is
\begin{equation}
\mbox{Tr}[M_t(q_0,p_0)]=  \sum_{j=0}^{[t/2]} (-1)^j \sum_i^{{odd\atop arrng.}} \prod_{k}^{{t-2j\atop factors}} \left[2- V^{\prime\prime}(q_{\{i,k\}})\right]
\end{equation}
where $[t/2]$ is the integer part of $t/2$ and the odd arrangements of the product of $t-2j$ factors is specified just ahead.  Also note that if $t$ is even, the final term (the one for which $j=t/2$) is $(-1)^{t/2}\times2$.  

There are two key properties required for the above  identity.  First, there must exist at least one column or row in the set of $\{M(q_i)\}$ both of whose elements are the same constants for all members of the set.  Second, the determinants of the set of $\{M(q_i)\}$ must all be equal to the same value (if this value is not equal to unity, then the terms in the equation above must be multiplied by its $j^{th}$ power) .  

It turns out there is a fairly simple rule for which products can contribute to a particular ``$j$-term''.   Imagine the $t$ integers $\{i\in [0,t-1]\}$ uniformly spaced on a ring.  Those are the ``locations'' that each factor in a product can occupy, so-to-speak, determined by its own subscript.  For a given time $t$ with $t-2j$ factors in the product, each arrangement that has exclusively odd-unit-spacings between occupied locations contributes once to that $j$-term.  All cyclically connected arrangements are automatically included in this construction, and thus the cyclic invariance of the trace is easy to see.  The number of such odd spacing arrangements ${\cal N}_A$ turns out to be 
\begin{equation}
\label{oddar}
{\cal N}_A = {t\over t-j}\left({t-j\atop j} \right)
\end{equation}
which is related to the summation and recursion relation
\begin{eqnarray}
\label{sumfu}
{\cal S}(t,j)&=& {1\over  j!}  {{\rm d}^j\over {\rm d}x^j} \left[\left(1+\sqrt{x}\right)^t+\left(1-\sqrt{x} \right)^t \Bigg]\right|_{x=1} \nonumber \\
&=& 2  \sum_{m=j}^{[t/2]} \left( {t\atop 2m}\right) \left( {m\atop j}\right)= 2^{t-2j}{t\over t-j}\left({t-j\atop j} \right) 
\end{eqnarray}
The combinatorials are related to an expansion of the functional form of eigenvalues raised to various powers of a two-by-two matrix.  The $j=0$ term is simply the product of all factors
\begin{equation}
\mbox{Tr}[M_t(q_0,p_0)]_{j=0} =  \prod_{i=0}^{t-1} \left[2- V^{\prime\prime}(q_i)\right]
\end{equation}
One could also label this term $\{1^t\}$ indicating that there are $t$ unit spacings.  Similarly, the $j=1$ term is always the cyclic sum $\{1^{t-3},3\}$ of which there are $t$ terms.  Thus,
\begin{equation}
\mbox{Tr}[M_t(q_0,p_0)]_{j=0+1} =  \prod_{i=0}^{t-1} \left[2- V^{\prime\prime}(q_i)\right] \times \left(1 - \sum_i^{cyclic} {1\over  \left[2- V^{\prime\prime}(q_i)\right] \left[2- V^{\prime\prime}(q_{i+1})\right] }  \right)
\end{equation}
where the last term has indices $(t-1,0)$.  Pulling the product of all factors out front emphasizes that if the $\{V^{\prime\prime}(q_i)\}$ are generally much larger than $2$, then for times not too long the $j>0$ terms can be thought of as corrections, each successive one roughly proportional to $t<(V^{\prime\prime})^{2}>^{-1}$ weaker that the previous term.  In some sense, the first non-trivial $j$-term comes for $t=6$ and $j=2$.   One expects $9$ double products of factors.  The allowed odd-spacing arrangements are $\{1,5\}$ and $\{3,3\}$.  There are $6$ terms related to the cyclic permutations of the $i(i+1)$ product  ($\{1,5\}$) and only $3$ terms for the $i(i+3)$ product ($\{3,3\}$) since it repeats itself after half a cycle, and thus $9$, the correct count of terms, is consistent with Eq.~(\ref{oddar}).
The above expression is the expansion of the determinant given by Bountis and Helleman~\cite{Bountis81} nearly 30 years ago.  Greene developed its small-$K$ expansion in his study of the behavior of KAM (Kolmogorov-Arnol'd-Moser) surfaces~\cite{Greene79}.

\vskip .4cm

{\bf Appendix B}

\vskip .4cm

It is more convenient to focus on the cumulants than the central moments as it simplifies the time dependence.  There is also the advantage of interpretation of the probability density.  For a Gaussian density, only the first two cumulants are non-zero.  For other densities, the higher cumulants describe shape deviations.  There is a third advantage, which comes into play when considering the relationship with quantum mechanics, but we do not elaborate here.  If the dynamical correlations vanish or are small enough to be ignored, then the expression for the cumulants can be simplified.  With these quantities, the cumulants have the same polynomial form, $P_n$, as with the central moments and a very simple dependence on $t$.
\begin{figure}[t]
\begin{center} 
\leavevmode 
\epsfxsize = 15.0cm 
\epsfbox{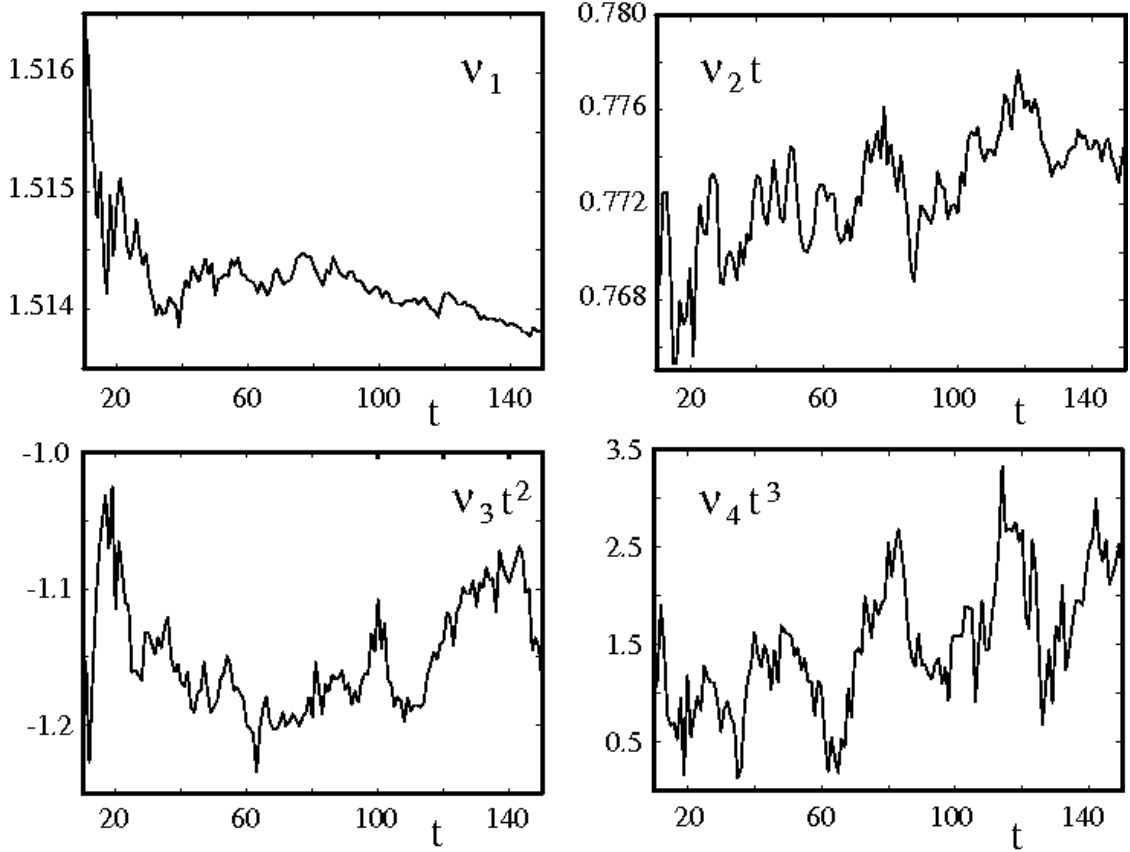} 
\end{center} 
\caption{The first four cumulants as a function of time.  The cumulants are multiplied by the time factors needed to make them independent of time according to the expressions leaving out dynamical correlations.  The kicking strength is 8.9633781407, which gives the prediction that the Lyapunov exponent (first cumulant) is equal to $1.513$.}
\label{lyap1}
\end{figure} 
The result is 
\begin{equation}
\label{scale}
\nu_n(t) = t^{1-n} P_n
\end{equation}
For example, 
\begin{eqnarray}
\nu_3(t) &=& {1\over t^2} {\cal X}_3 \nonumber \\
\nu_4(t) &=& {1\over t^3}\left( {\cal X}_4-3{\cal X}_2^2\right)  
\end{eqnarray}
An amusing, known consequence of the time dependence in Eq.~(\ref{scale}) is that  there is a sense in which the distribution collapses to a $\delta$-function density as $t\rightarrow \infty$ as expected (every trajectory has the same Lyapunov exponent), and the time-dependence of the reduced cumulants goes as
\begin{equation}
\tilde \nu_n(t) = {\nu_n(t)\over \nu_2^{n/2}(t)} \propto t^{1-n/2}
\end{equation}
which for $n\ge 3 \rightarrow 0$ as $t\rightarrow \infty$.  In other words, the density slowly approaches a Gaussian form.  Nevertheless, the $\delta$-function limit is approached slowly enough that the width of the probability density for $|\mbox{T}r(M_t)|$ grows without bound as $t$ increases, and it is not well approximated by a lognormal density.   An example of the time dependence of the first four cumulants for a kicked rotor can be seen in Fig.~(\ref{lyap1}).  
It illustrates that the time dependence is to within fluctuations as predicted for $K$ values at which dynamical correlations are not observed.

\section*{Acknowledgments}

The authors gratefully acknowledge important discussions with Denis Ullmo, Alfredo Ozorio de Almeida, and Shmuel Fishman.   They also gratefully acknowledge Roland Ketzmerick for finding kicking strength of the double saddle-node bifurcation, and S. T. gratefully acknowledges support from the US National Science Foundation grant PHY-0555301.

\bibliography{classicalchaos,quantumchaos,rmt}

\end{document}